\documentclass[12pt]{article}    
\pdfoutput=1

\usepackage[vmargin={1.25in, 1.25in},hmargin={1.15in, 1.15in}]{geometry}

\usepackage{latexsym}
\usepackage{amssymb}
\usepackage{amsmath}
\usepackage{amsthm}
\usepackage{mathtools}
\usepackage{amsfonts}
\usepackage{graphicx}
\usepackage{enumerate}
\usepackage{color}
	\definecolor{MyDarkBlue}{rgb}{0,0.08,0.45}
\usepackage{hyperref}
	\hypersetup{linkcolor=MyDarkBlue, citecolor=MyDarkBlue, colorlinks=true}
\usepackage{setspace}
\usepackage[longnamesfirst]{natbib}

\usepackage{hypernat}
\newtheorem{prop}{Proposition}
\newtheorem{lemma}{Lemma}
\newtheorem{corollary}{Corollary}

\theoremstyle{definition}

\newcommand{\finexhere}{\begingroup \let\mathqed\math@qedhere
    \let\qed@elt\setQED@elt \blacktriangle@stack\relax\relax \endgroup}

\newcommand{\bE}{\mathbb{E}}

\newcommand{\bR}{\mathbb{R}}

\newcommand{\supp}{\mathrm{supp\,}}

\title{\vspace*{-3em}Variational Bayes and non-Bayesian  Updating}
\author{Tomasz Strzalecki\thanks{Department of Economics, Harvard University. Date: \today. I thank Ryota Iijima for alterting me to a very close result of \cite{Dominiak2023inertial}, which I was unaware of when writing this note.}}
\date{}

\begin{document}
	
\maketitle

\begin{abstract}
	I show how variational Bayes can be used as a microfoundation for a popular model of non-Bayesian updating.
\end{abstract}

\section{Bayesian Updating}
Let $S$ be a finite state space (parameter space) and $p\in \Delta(S)$ be the prior of the agent. The agent gets information: let $X$ (finite) be the set of possible signal realizations (messages). Let $f:S\to\Delta(X)$ be the \emph{likelihood} function (a theorist may call $f$ a Blackwell experiment). Let $q_x\in \Delta(S)$ be the agent's belief upon observing data $x$. As we all know, the Bayes rule says that
\begin{equation}
q_x(s)=\frac{f(x|s) p(s)}{\sum_{s'\in S}f(x|s') p(s')} \label{eq:Bayes}
\end{equation}
The denominator, the marginal on $x$, is often hard to compute in practice when $X$ is continuously distributed and the sum is an integral. In the next note I will talk about one particular computational approach to this problem, called \emph{variational Bayes}. For now, to get rid of the denominator, define $\propto$  as ``equal up to a multiplicative factor'' and write \eqref{eq:Bayes} as
\begin{equation}
q(s) \propto p(s) f(x|s).\label{eq:Bayes-prop}
\end{equation}

\section{Non-Bayesian Updating}
Researchers in psychology have found that experimental subjects deviate from the Bayes rule in systematic ways.\footnote{\cite{edwards1964man}; \cite{phillips1966conservatism}; \cite{tversky1971belief}; \cite{kahneman1972subjective}.} Such deviations are classified as normatively incorrect (biases, errors, etc.). \cite{Benjamin19} reviews this evidence and uses the following two-parameter generalization of \eqref{eq:Bayes-prop} as a descriptive model to categorize various errors in probability updating. \cite{AR} also use this model to obtain a similar classification. 

For lack of better terminology I will call this the \emph{exponential updating rule}
\begin{equation}
q_x(s)\propto  \big[p(s)\big]^\alpha \big[f(x|s)\big]^\beta,\label{eq:expo}	
\end{equation}
where $\alpha, \beta >0$. The exponent $\alpha$ measures how much the agent over/under weighs the prior and the exponent $\beta$ measures how much they overweigh the evidence. In particular, $\alpha<1$ is classified as underinference (or conservatism), $\alpha>1$ as  overinference, $\beta<1$ as base-rate neglect, and $\beta>1$ as confirmation bias. As noticed by \cite{Grether80} if the researcher runs a log linear regression of posteriors over priors and likelihoods (without an intercept), then they are de facto estimating a exponential updating rule.

Interestingly, the case of $\alpha=1$, $\beta<1$ has been proposed as a normative model in Bayesian statistics\footnote{ \cite{Grunwald12}; \cite{BissiriHolmesWalker16}; \cite{Bhattacharya19}.} and machine learning.\footnote{\cite{Huang2018improving}; \cite{higgins2017beta}; \cite{Pepe}.} Thus, what would be characterized by a behavioral economist as an error in probability updating, to a statistician may be an appealing way to guard against model misspecification. This way of modeling concern for misspecification is distinct from the economics literature, where the focus is not directly on misspecification of the likelihood function $f$, but on misspecification of the of predictive posterior distribution.\footnote{\cite{HS}; \cite{cerreiavioglio2022making}; \cite{lanzani2002dynamic}.}

\section{Variational Bayes}
Consider now a different approach, where instead of an explicit formula for updating beliefs, the agent is choosing her beliefs to minimize some loss function. Fix the prior $p\in \Delta(S)$. Recall the \emph{relative entropy} (Kullback-Leibler divergence) $D:\Delta(S)\times\Delta(S)\to [0, +\infty]$, given by 
$$D(q\parallel p)=\sum_{s\in S}q(s)\log\frac{q(s)}{p(s)}$$
when $q\ll p$ and $+\infty$ otherwise.\footnote{We use the convention that $0\log\frac{0}{c}=0$ for any $c\geq 0$. For any $q\in \Delta(S)$ we write $q\ll p$ whenever $p(s)=0$ implies $q(s)=0$ (that is $q$ does not invent new states that $p$ does not believe in). The Bayesian posterior has this property and any alternative updated belief considered by the agent will also have it.} Relative entropy is a notion of ``distance'' between $q$ and $p$. If we solve
$$\min_{q\in \Delta(S)} D(q\parallel p),$$
we will get $q=p$.\footnote{This is known as the ``information inequality,'' Theorem 2.6.3. of \cite{CoverThomas91}.}

So far we have gotten nowhere in terms of updating. We started with a prior and our procedure ended up with a prior, so we have not updated at all. This is because we have not used the data yet. Consider now a different optimization problem:
$$\max_{q\in \Delta(S)}   \sum_{s\in S}  q(s) \log f(x|s).$$
Here we are choosing $q$ to best fit the data. The solution will concentrate all its mass on the states that lead to maximum likelihood $f (x|s)$. Maximum likelihood estimation is the classical approach to statistics.

Now the key point is that we get Bayesian updating if we combine the two problems: 
\begin{equation}
\min_{q\in \Delta(S)}  D(q\parallel p) -  \sum_{s\in S}  q(s) \log f(x|s).	\label{eq:VB}
\end{equation}

\begin{prop}\label{prop:VB}
	The solution to \eqref{eq:VB} is unique and equal to the Bayesian posterior \eqref{eq:Bayes}.
\end{prop}

Thus, Bayesian updating can be understood as striking a balance between being close to the prior and being close to the evidence.  This way of looking at things is known as \emph{variational Bayes} (\citealt{peterson1987mean}; \citealt{hinton1993keeping}; \citealt{Bleietal17}).\footnote{The adjective ``variational'' is here because we are solving an optimization problem (by calculus of variations when $S$ is infinite).} To be more precise: It is as hard to solve \eqref{eq:VB} as it is to compute \eqref{eq:Bayes} (since we are getting the same answer). The computational advantage in variational Bayes comes from \emph{regularizing} the problem, which means adding a set of constraints (e.g.,  $q$ is a product measure on $S=\bR^d$) or adding additional terms to the objective function. I will abstract from this and focus on the exact equivalence as in Proposition~\ref{prop:VB}.

\section{Model misspecification}
So far we understood that Bayesians are striking a balance between priors and data. But why equal weights? Consider this problem: 
\begin{equation}
\min_{q\in \Delta(S)}  D(q\parallel p) -  \lambda \sum_{s\in S}  q(s) \log f(x|s),	\label{eq:lambda}
\end{equation}
where $\lambda>0$ is a parameter. 

Here $\lambda$ measures the attitude of our agent toward model misspecification. If the agent has full confidence in the model, she sets  $\lambda=1$ and gets Bayes. But with imperfect confidence she might want to set $\lambda<1$. As mentioned before, this has been proposed as a normative model in Bayesian statistics. On the other extreme is  $\lambda=\infty$, which as we saw above, corresponds to maximum likelihood estimation, a procedure in classical statistics.

\begin{prop}\label{prop:WB}
	The solution to \eqref{eq:lambda} satisfies
$$q_m(s)\propto p(s)\big[f(x|s)\big]^\lambda.$$
\end{prop}

This is exponential updating with $\alpha=1$ and $\beta=\lambda$. Thus, taking Bayes as a point of reference, $\lambda<1$ leads to underweighting the evidence, whereas $\lambda>1$ leads to overweighting the evidence.

Overweighting the evidence seems similar to underweighting the prior. This is true at the level of ``preferences over beliefs'' represented by \eqref{eq:lambda}: increasing $\lambda$ puts more weight on the evidence and less relative weight on the prior. However, at the level of the updating rule, this is not exactly true: in the exponential updating formula there is a separate parameter, $\alpha$, which measures how much we over/underweight the prior. What force does this correspond to at the level of ``preferences over beliefs?'' I explore this in the next section.

\section{Attitudes toward uncertainty}
Recall that the entropy of $q$ is defined by 
$$H(q):=-\sum_{s\in S} q(s)\log q(s),$$ 
with the convention $0\log 0=0$. The function $H$ measures how chaotic or uncertain the distribution $q$ is. 
 
Suppose now that our decision maker has a taste for entropy and solves:
\begin{equation}
	\min_{q\in \Delta(S)} D(q\parallel p) - \lambda \sum_{s\in S} q(s) \log f(x|s) + \mu H(q).	\label{eq:hybrid}
\end{equation}
for some parameters $\lambda>0, \mu\in \bR$.

The case $\mu<0$ (uncertainty loving) was studied by \cite{Rava} under the assumption that $\lambda=1$ (no model misspecification).\footnote{Formally, their agent minimizes $D(q\parallel q^{Bayes}_x) + \mu H(q)$ for some $\mu<0$.} This agent dislikes having beliefs that are close to certainty. A related notion of extreme-belief aversion was studied by \cite{benjamin2016model}. On the other hand, if the agent prefers to have simple theories of the world and holding complex beliefs is more costly \citep{enke2023cognitive}, then we have $\mu>0$ (uncertainty aversion).

The next proposition shows that such an agent either follows the exponential updating rule (if they are not too averse toward uncertainty) or puts all posterior probability on one state (if they dislike uncertainty too much). 

\begin{prop}\label{prop:HB}
	Suppose that $\lambda>0$. If $\mu< 1$ the solution to \eqref{eq:hybrid} is the exponential updating rule with $\alpha:=\frac{1}{1-\mu}$ and $\beta:=\frac{\lambda}{1-\mu}$. If $\mu\geq 1$, then the solution puts all probability on states that maximize  $p(s)\left(f(x|s)\right)^\lambda$.
\end{prop}

Behavior ``flips'' around $\mu=1$. The objective function is convex as long as $\mu< 1$. For $\mu\geq 1$ the objective function is concave and the minimum value is attained at the extreme points of the simplex.  The case of $\mu\geq 1, \lambda=1$ corresponds to Maximum a Posteriori (MAP) estimation in Bayesian statistics, i.e., the mode of the posterior distribution \citep{robert2007bayesian}.

\begin{corollary}\label{cor:HB}
	Any exponential updating rule $(\alpha, \beta)$ is a solution to \eqref{eq:hybrid} with 
	\begin{equation}
	\lambda=\frac{\beta}{\alpha} \text{ and } \mu=1-\frac{1}{\alpha}. \label{eq:params-trans}	
	\end{equation}	
\end{corollary}




%
%


\section{Related Literature}

\cite{Dominiak2023inertial} study a model of updating where the agent has a state space $\Omega$ and receives information in the form of $E\subseteq \Omega$. Their agent chooses their posterior to minimize a general notion of distance between their posterior and their prior, subject to the constraint that the posterior puts probability one on event $E$. They show that in the case when $\Omega=S\times X$ and $E=S\times \{x\}$, a specific cost function leads to the exponential updating rule considered here. My Proposition \ref{prop:HB} accomplishes the same, but with the added value of decomposing the preferences into attitudes toward model misspecification and toward uncertainty. 

\cite{zhao2022pseudo} studies a related model of updating upon \emph{qualitative} information, where the update is a solution to minimizing relative entropy. His result is an analogue of Proposition \ref{prop:VB}. 



\cite{Geffroy} study non-Bayesian persuasion where the receiver uses a updating rule such that the updated belief is a function only of the prior belief and the correct Bayesian posterior. They show that exponential discounting satisfies this property.

\cite{cripps2018divisible} studies a general class of updating rules where the order of signal matters. The $\alpha=1$ case of exponential updating satisfies this property, but for $\alpha\neq 1$ the order of signals matters.

\citealt{EpsteinNoorSandroni} study a model where the updated belief is an affine combination of the prior and the Bayesian posterior. In this note the mixing operates in the log-probability space, not the probability space. 

In the literature on motivated beliefs agents optimally choose their beliefs subject to some constraints or costs (\citealt{BrunnermeierParker05}; \citealt{BenabouTirole16}). From this point of view, Proposition \ref{prop:VB} identifies the objective function that leads to the benchmark case of Bayes updating. 

  
\appendix

\section{Proofs}
\subsection{A Variational Lemma}

The following variational formula is  well-known to economists (\citealt{Stahl90}, \citealt{APT92}, \citealt{FL98}, \citealt{HS}). It also holds for infinite $S$, but I will limit my discussion to finite spaces to avoid measurability issues.

\begin{lemma}\label{lem:DE}
	For any function $g:S\to \bR$ the minimum in
		$$\min_{q\in \Delta(S)} \bE_q g - H(q)$$
		is attained uniquely by 
		$$q(s)\propto \exp(-g(s)).$$
\end{lemma}

\subsection{Proof of Proposition \ref{prop:VB}}
Fix the prior $p\in \Delta(S)$. Given that $D(p\parallel q)$ is infinite for $q\ll p$, the minimum will be attained at $q\in \Delta(S^*)$ where $S^*:=\supp(p)$. Set $g(s):=-\left[\log p(s) + \log f(x|s)\right]$. 

\subsection{Proof of Proposition \ref{prop:WB}}
Set $g(s):=-\left[\log p(s) + \lambda \log f(x|s)\right]$. 

\subsection{Proof of Proposition \ref{prop:HB}}
When $\mu\leq 1$, set $g(s):=-\frac{1}{1-\mu}\left[\log p(s) + \lambda \log f(x|s)\right]$. When $\mu\geq 1$,  we are minimizing a concave function, so the minimum value will be obtained at some extreme point. Which one? The Dirac on state $s\in S$ that maximizes $p(s)\left(f(x|s)\right)^\lambda$. If there are multiple such states, then any mixture between them is also a solution.

\begin{spacing}{1.25}
\bibliographystyle{econometrica}
\bibliography{../bib}
\end{spacing}

\end{document}